\documentclass[cameraready]{Interspeech}

\title{Improving Zero-Shot Phonetic Classification \\through Language-Agnostic Articulatory Features}

\author[affiliation={1}, orcid=0009-0009-5845-8675]{Ryo}{Magoshi}
\author[affiliation={1,2}, orcid=0009-0001-4880-2059]{Jaeyoung}{Lee}
\author[affiliation={1}, orcid=0009-0006-6883-1603]{Shinsuke}{Sakai}
\author[affiliation={1}, orcid=0000-0002-2686-2296]{Tatsuya}{Kawahara}

\address{
    $^1$ Graduate School of Informatics, Kyoto University, Japan \\
    $^2$ NTT, Inc., Japan
}

\email{magoshi@sap.ist.i.kyoto-u.ac.jp, jalpr16@gmail.com, sakai@sap.ist.i.kyoto-u.ac.jp, kawahara@i.kyoto-u.ac.jp}

\keywords{zero-shot phonetic recognition, IPA transcription, articulatory features}

\usepackage{comment}
\usepackage{amsmath}
\usepackage{tipa}
\usepackage{multirow}
\usepackage{float}
\usepackage{here}
\usepackage{tikz}
\usetikzlibrary{patterns,positioning,arrows.meta,calc,decorations.pathreplacing}
\usepackage{stfloats}
\usepackage{cite}

\begin{document}

\maketitle
\begin{abstract}
Recent Phonetic Foundation Models (PFMs) for Speech-to-IPA transcription rely on Grapheme-to-Phoneme (G2P) labels, but the phoneme labels are not necessarily phonetically faithful.
To investigate this issue, we evaluate zero-shot phonetic classification on Chinese aspiration and Japanese moraic nasals.
A PFM trained on G2P-labeled data excluding these two languages yields poor accuracy on both tasks, showing that multilingual coverage with discrete IPA tokens is not sufficient for unseen settings.
To overcome this limitation, we propose a classification method based on continuous Articulatory Feature (AF) vectors extracted from each frame.
This AF-based approach outperforms discrete token-based methods, particularly for rare phones.
We further show that it is crucial to adopt the optimal temporal aggregation of AF vectors for the target distinction: single-frame classification is best for aspiration, while segmental classification substantially improves nasal classification.

\end{abstract}

\section{Introduction}

Automatic transcription of speech into the International Phonetic Alphabet (IPA), or Speech-to-IPA, is an important task.
In principle, the IPA's universality makes it possible to represent all human speech sounds within a single framework.
Thus, Speech-to-IPA is designed to transcribe acoustic reality directly beyond the limits of language-specific phonemic inventories.
This objective distinguishes it from standard ASR, which is limited by the orthographic rules of specific writing systems, and opens up several applications.
First, it is essential for the analysis of atypical speech, such as the pronunciation of L2 learners~\cite{zhao18b_interspeech,shahin_2024} and the evaluation of pathological speech~\cite{ipapathol}.
Second, it contributes to transcription of endangered languages that lack a written form, aiding in cultural preservation~\cite{foley18_sltu,shi-etal-2021-leveraging}.
Third, sharing an IPA token set across languages facilitates knowledge transfer from high-resource to low-resource languages~\cite{allosaurus,xu22b_interspeech,yen_boosting,jaeyoung_ipa_asr,magoshi25_interspeech,mimura25_interspeech}.

Speech-to-IPA research based on deep learning has progressed rapidly.
This progress has led to the emergence of Phonetic Foundation Models (PFMs) that claim universality through scaling~\cite{zhu-etal-2025-zipa,li2026powsmphoneticopenwhisperstyle}.
However, these models encounter a critical issue regarding the phonetic validity of their training data.
Most current datasets for Speech-to-IPA rely on Grapheme-to-Phoneme (G2P) conversion to generate labels.
As a language-dependent processing, G2P maps graphemes to IPA symbols without observing the actual acoustic signal.
This process inherently introduces conversion errors and inconsistent IPA notations across languages, making the labels phonemic rather than phonetic.
Consequently, acoustic differences that helps distinguishing phonemes in one language may be treated as ignorable variations in another.
For example, a G2P system for a language that does not distinguish aspiration will ignore these acoustic nuances and assign an identical label to different phones.
These characteristics conflict with the goal of Speech-to-IPA, preventing the model from acquiring truly universal articulatory recognition.

While existing PFMs claim zero-shot performance through multilingual training~\cite{zhu-etal-2025-zipa,li2026powsmphoneticopenwhisperstyle}, these evaluations often rely on unreliable G2P-derived data and language-specific training.
To investigate this issue, this study first evaluates Speech-to-IPA models in a zero-shot setting.
We select evaluation tasks where rule-based G2P produces reliable labels for the target contrasts, and show that current PFMs perform poorly when discriminating unseen phones.
These findings demonstrate that multilingual coverage alone is not sufficient for universal recognition in zero-shot settings, even if the target IPA tokens are included in the training data and are shared across languages.

To overcome the limitations, we propose a classification framework based on 24-dimensional Articulatory Features (AF) estimated by a large-scale self-supervised model.
The Articulatory Feature Classification Module (AFCM) is jointly trained with the IPA tokens prediction and provides a phonetic-level representation.
Therefore, it realizes language-agnostic classification by examining relevant speech frames.

This study makes two contributions. First, we reveal the limitations of current multilingual training approaches through zero-shot classification on distinct phones. 
Second, we propose an AF-based classification framework tailored to the acoustic and phonetic nature of the target distinction.

\begin{figure*}[t]
\centering
\includegraphics[width=\textwidth]{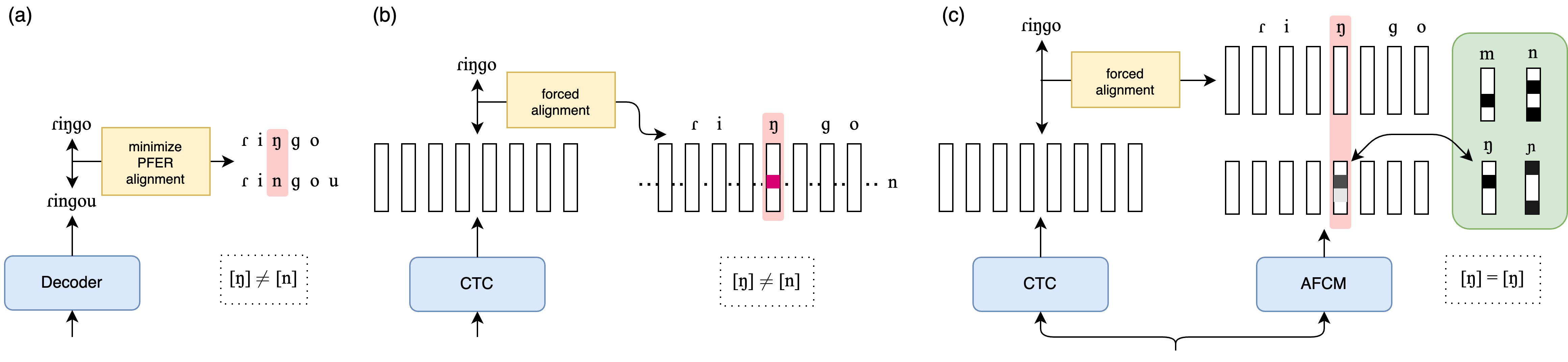}
\caption{Overview of three classification methods, illustrated on a Japanese word ``ringo'' [\textipa{RiNgo}] (apple) for a nasal classification task. (a)~Decoder-based classification aligns the decoded IPA sequence with the reference via PFER minimization. (b)~CTC-based classification takes $\arg\max$ of the frame-level posteriors at the peak frame $t^*$ identified by forced alignment. (c)~AF-based classification (proposed) computes the L1 distance between the continuous AF vector at $t^*$ and PanPhon template vectors, selecting the closest match.}
\label{fig:method}
\vspace{-0.5em}
\end{figure*}

\section{Related Work}

Automatic transcription of speech into phonetic or phonemic representations can be categorized into two approaches.

The first approach employs either the IPA, or language-specific symbols such as Hiragana for Japanese or Pinyin for Chinese within a specific linguistic framework~\cite{rezackova21_interspeech,ohnaka25_interspeech,mak25_interspeech,hu25_interspeech}. 
These models can be accurate enough to serve as reliable front-ends for text-to-speech systems~\cite{ohnaka25_interspeech,mak25_interspeech,hu25_interspeech}. 
Consequently, they remain constrained by language-specific phonology and lack the capacity for universal transcription across diverse languages.
Even when the IPA is utilized, selection of the symbol set is typically biased toward the phonemic inventory of the target language. 

The second approach utilizes a common symbol set, primarily the IPA, to achieve language-universal phonetic recognition~\cite{taguchi23_interspeech,magoshi25_interspeech,zhu-etal-2025-zipa,li2026powsmphoneticopenwhisperstyle,suchardt-etal-2025-stipa}.
Speech-conditioned G2P systems, which refine grapheme-to-phoneme conversion using acoustic information, also fall into this category~\cite{route_multimodal_2019,ribeiro23b_interspeech,g2pu}.
While this approach can be applied to multilingual settings, it often suffers from degraded accuracy due to inconsistencies and errors in the underlying IPA labels.
The primary bottleneck is that most large-scale datasets rely on G2P conversion tools rather than manual transcription, introducing the following issues:

\begin{enumerate}
\item Acoustic-Agnostic Labeling: Because G2P operates solely on orthographic input, the resulting labels represent abstract phonemes and fail to capture the natural phonetic variations present in actual speech.
\item Notational Divergence: Despite the IPA's conceptual universality, notation conventions and symbol subsets vary significantly across languages and datasets, resulting in a lack of inter-dataset consistency.
\item Algorithmic Errors: G2P systems, whether rule-based~\cite{mortensen-etal-2018-epitran,phonemizer_espeak,matogawa-etal-2024-japanese}, 
statistical~\cite{BISANI2008434,deri-knight-2016-grapheme,phonetisaurus} 
or neural~\cite{g2p_lstm,yolchuyeva19_interspeech,park20c_interspeech,zhu22_interspeech,li-etal-2022-zero,llm_g2p}, are prone to conversion errors that propagate into training labels.
\end{enumerate}

These factors hamper phonetic faithfulness, hindering the development of models capable of universal recognition.
On the other hand, AFs are phonetic descriptors that characterize each IPA symbol along dimensions such as nasality, voicing, and place of articulation, providing a continuous-valued and language-agnostic representation of speech sounds.
To mitigate the limitations of discrete phonetic tokens, several studies have explored the use of AFs as an auxiliary or alternative representation for speech recognition.

Li et al.~\cite{li19j_interspeech} utilized AFs as intermediate representations for grapheme-level output. 
However, their AFs are discrete and language-specific, precluding zero-shot application.
Yen et al.~\cite{yen_boosting} employed language-universal speech attribute constraints to improve multilingual phoneme recognition, demonstrating that AF-based regularization improves performance.
Both approaches used AFs with only two dimensions, place and manner of articulation, which limits their phonetic expressiveness.

Magoshi et al.~\cite{magoshi25_interspeech} adopted the AF system defined by PanPhon~\cite{mortensen-etal-2016-panphon}, which converts each IPA token into a 24-dimensional AF vector whose dimensions encode phonetic properties such as syllabic, sonorant, nasal, and continuant.
Each dimension takes a value of $+1$ (present), $-1$ (absent), or $0$ (not applicable), covering IPA symbols including those with diacritics such as aspirated stops.
This representation is richer than the two-dimensional characterizations used in prior work.
They proposed a frame-level AF classification method as an auxiliary task within a CTC-based Speech-to-IPA model, producing continuous AF vectors alongside discrete IPA posteriors.
Our work builds upon~\cite{magoshi25_interspeech} but uses the AF vectors directly for phonetic classification, rather than relying on discrete IPA token outputs.

\section{Methods}

We compare two Speech-to-IPA models (\S\ref{sec:models}) and design three complementary classification methods to assess their zero-shot phonetic discrimination capability (\S\ref{sec:classification}).

\subsection{Models}\label{sec:models}

POWSM~\cite{li2026powsmphoneticopenwhisperstyle} is a hybrid CTC/Attention model~\cite{hybrid_ctc_attn}.
An Attention-based Encoder-Decoder (AED) autoregressively generates an IPA token sequence, while a Connectionist Temporal Classification (CTC)~\cite{ctc} branch provides frame-level posterior distributions over the IPA token set.
This architecture is representative of recent large-scale PFMs~\cite{li2026powsmphoneticopenwhisperstyle}.

XLS-R + AFCM is a CTC-based model that uses XLS-R~\cite{babu22_interspeech} as a pre-trained encoder, 
augmented with an Articulatory Feature Classification Module (AFCM)~\cite{magoshi25_interspeech} which estimates a 24-dimensional AF vector at each frame in addition to the standard CTC posteriors.
We propose using this AF output for zero-shot phonetic classification.
The AFCM is trained with cross-entropy loss to predict the PanPhon~\cite{mortensen-etal-2016-panphon} AF vector corresponding to each frame's IPA token. 
The resulting logits constitute the continuous AF vectors used for classification.
It is jointly trained with CTC-based IPA token recognition. 
Although the AF labels inherit errors from the IPA labels, the AFCM learns distributed phonetic properties rather than discrete tokens, making it more robust to individual labeling errors.


\subsection{Classification Methods}\label{sec:classification}

To assess whether each model correctly identifies the target IPA symbol, we design three classification methods (Figure~\ref{fig:method}) that share a common two-step procedure: (1) locate the segment in which the target IPA is expected to occur, and (2) examine the model's output within that segment to determine correctness.
Decoder-based and CTC-based classifications serve as baselines that operate on discrete IPA token outputs.

Decoder-based classification aligns the decoded IPA sequence $\hat{\mathbf{Y}} = (\hat{y}_1, \ldots, \hat{y}_N)$ to the reference sequence $\mathbf{Y}^* = (y^*_1, \ldots, y^*_M)$ via $\pi^* = \arg\min_{\pi} \mathrm{PFER}(\hat{\mathbf{Y}}, \mathbf{Y}^*; \pi)$, minimizing the Phoneme-Feature Error Rate (PFER)~\cite{taguchi23_interspeech}.
Here, $\pi$ denotes an alignment mapping from the reference indices to the decoded indices.
For the target position $k$, the prediction is deemed correct if the aligned token $\hat{y}_{\pi^*(k)}$ matches the ground truth $y^*_k$.
If the alignment maps no decoded token at the position $k$ (i.e., deletion), the prediction is treated as incorrect.

CTC-based classification utilizes the CTC log-posterior probabilities $\log p(c \mid \mathbf{x}_t)$, where $c$ denotes an IPA token and $\mathbf{x}_t$ is the encoder output at frame $t$. Given the interval $(t_s, t_e)$ of the target segment obtained by forced alignment, the peak frame $t^*$ is identified as:
\begin{equation}
    t^* = \arg\max_{t_s \leq t \leq t_e} \log p(c^{\mathrm{ref}} \mid \mathbf{x}_t)
\end{equation}
where $c^{\mathrm{ref}}$ is the reference token. The estimated token is then $\hat{c} = \arg\max_{c} \log p(c \mid \mathbf{x}_{t^*})$.

Building on the CTC framework, we propose AF-based classification, which replaces discrete token posteriors with continuous AFs. Let $\boldsymbol{a}_t$ denote the AFCM output at frame $t$, and let $t^*$ be the peak frame obtained as in CTC. 
We estimate the phonetic identity by computing the L1 distance between the AF vector $\boldsymbol{a}_{t^*}$ and the template vector derived from each candidate category $c \in \mathcal{C}$:
\begin{equation}
    \hat{c} = \arg\min_{c \in \mathcal{C}} \sum_{i \in \mathcal{D}} |a_{t^*,i} - f_{c,i}|
\end{equation}
where $\mathbf{f}_c$ is the PanPhon AF template vector for IPA token $c$, $\mathcal{C}$ is the set of candidate IPA tokens, and $\mathcal{D}$ is the set of AF dimensions that are discriminative among $\mathcal{C}$.
For example, the four Japanese nasals [m], [n], [\textipa{N}], [\textipa{\textltailn}] share identical values on all AF dimensions except five place features (anterior, coronal, labial, high, and back), yielding $|\mathcal{D}| = 5$.

For both CTC-based and AF-based methods, we compare two temporal aggregation methods. The \textit{single-frame} classification directly uses the representation at the peak frame $t^*$. The \textit{segmental} classification instead averages the frame-level representations from $t^*$ to the onset of the subsequent token:
\begin{equation}
    \bar{\mathbf{v}} = \frac{1}{T_{\mathrm{next}} - t^*} \sum_{t=t^*}^{T_{\mathrm{next}}-1} \mathbf{v}_t
\end{equation}
where $T_{\mathrm{next}}$ is the onset frame of the subsequent IPA token and $\mathbf{v}_t$ denotes the CTC log-posterior vector or the AF vector $\mathbf{a}_t$. 
Then, respective classification method is applied to $\bar{\mathbf{v}}$ in place of the single-frame representation.

\section{Experiments}

\subsection{Experimental Settings}

The two models are trained on the Common Voice~\cite{common_voice} and FLEURS~\cite{fleurs} portions of IPAPack++~\cite{zhu-etal-2025-zipa}, which provides G2P-derived IPA labels for multiple speech corpora.
After excluding Chinese and Japanese, the training set covers 78 languages and approximately 3,000 hours.
IPA sequences in the training data are tokenized using PanPhon~\cite{mortensen-etal-2016-panphon}, yielding a vocabulary of 280 tokens.
Table~\ref{tab:lang_family} shows the language family distribution of the training data.
The ``Other'' category groups six families each consisting of five or fewer languages: Austronesian, Uralic, Mongolic, Kartvelian, Constructed (Interlingua), and one language isolate (Basque).

\begin{table}[t]
\centering
\caption{Language family distribution of the training data.}
\label{tab:lang_family}
\footnotesize
\begin{tabular}{lrr}
\hline
Language family & \# Lang. & Hours \\
\hline
Indo-European   & 40 & 2115.0 \\
Atlantic-Congo  &  8 &  568.2 \\
Dravidian       &  3 &   85.4 \\
Turkic          &  6 &   79.9 \\
Kra-Dai         &  2 &   46.2 \\
Sino-Tibetan    &  3 &   35.5 \\
Afro-Asiatic    &  3 &   16.4 \\
Austroasiatic   &  1 &    8.2 \\
Other (6 families) & 12 & 45.1 \\
\hline
Total           & 78 & 3000 \\
\hline
\end{tabular}
\vspace{-0.5em}
\end{table}

\begin{table}[t]
\centering
\caption{Distribution of target phones in the training data. \\ ``Ratio” denotes the proportion of aspirated tokens within each pair of voiceless stops (top) and the proportion of each phone in the nasal group (bottom).}
\label{tab:phone_dist}
\footnotesize
\begin{tabular}{lrr}
\hline
Phone & Count & Ratio \\
\midrule\midrule
{[p\textsuperscript{h}]} & 192k & \multirow{2}{*}{9.6\%} \\
{[p]} & 1.80M & \\
\cmidrule{1-3}
{[t\textsuperscript{h}]} & 264k & \multirow{2}{*}{4.9\%} \\
{[t]} & 5.12M & \\
\cmidrule{1-3}
{[k\textsuperscript{h}]} & 267k & \multirow{2}{*}{7.7\%} \\
{[k]} & 3.22M & \\
\midrule\midrule
{[m]} & 3.06M & 31.6\% \\
{[n]} & 5.63M & 58.3\% \\
{[\textipa{N}]} & 722k & 7.5\% \\
{[\textipa{\textltailn}]} & 253k & 2.6\% \\
\hline
\end{tabular}
\vspace{-1.2em}
\end{table}

POWSM follows the architecture of~\cite{li2026powsmphoneticopenwhisperstyle} but is trained from scratch only with the IPA tokens in our training set, resulting in 252M parameters due to the reduced vocabulary size.
\footnote{The original POWSM is not used because its training data includes Chinese and Japanese, which are used for the zero-shot evaluation.}
XLS-R + AFCM combines a pre-trained XLS-R encoder with a CTC head and AFCM~\cite{magoshi25_interspeech}, totaling 318M parameters.
Both models are trained with AdamW~\cite{Loshchilov2017DecoupledWD} at a learning rate of $5 \times 10^{-4}$, with 10k warmup steps and a batch size of 10 minutes.

We design two zero-shot tasks with Chinese and Japanese, which are excluded from the training dataset.
The first task classifies Chinese voiceless stops ([p\textsuperscript{h}]/[p], [t\textsuperscript{h}]/[t], [k\textsuperscript{h}]/[k]) as aspirated or unaspirated, evaluated on the Chinese FLEURS test set from IPAPack++ (652 utterances, 2.0 hours).
The second task classifies Japanese moraic nasals ([m], [n], [\textipa{N}], [\textipa{\textltailn}]), evaluated on the CSJ~\cite{csj} eval2 and eval3 (586 utterances, 52 minutes) with human-verified IPA labels~\cite{matogawa-etal-2024-japanese}.
In both cases, a rule-based G2P produces faithful labels: Chinese aspiration is a phonemic contrast and the Japanese nasals are conditional allophones which are predictable from the following consonant~\cite{vance2008sounds}.
In AF-based classification, the PanPhon AF template vectors $\mathbf{f}_c$ are binarized from $\{-1,+1\}$ to $\{0,1\}$ to match the binary output of the AFCM.
For evaluation, we report balanced accuracy, defined as the macro-average of per-IPA recall.

Because phone frequencies in natural language are inherently imbalanced, the training data reflects this skew.
Table~\ref{tab:phone_dist} shows the distribution of the target phones in the training data.
Aspirated stops account for fewer than 10\% of the voiceless stop tokens at every place of articulation, and [\textipa{\textltailn}] constitutes only 2.6\% of nasal tokens.

\begin{table*}[tb!]
\caption{Zero-shot phonetic classification results. (a) Three binary classifications of Chinese voiceless stops (aspirated vs.\ unaspirated) at each place of articulation. (b) A single four-way classification of Japanese nasals. All values are recalls in percentages and balanced accuracy is the macro-average of per-IPA recall. ``AF segmental'' denotes segmental AF classification.}
\label{tab:results}
\centering
\textit{(a) Chinese aspiration: binary classification per place of articulation}
\vspace{0.2em}

\begin{tabular}{llrcc!{\vrule width 0.3pt}cc!{\vrule width 0.3pt}ccc}
\hline
Model & \#Params & Method & {[p\textsuperscript{h}]} & {[p]} & {[t\textsuperscript{h}]} & {[t]} & {[k\textsuperscript{h}]} & {[k]} & Balanced Acc. \\
\midrule\midrule
\multirow{3}{*}{POWSM} & \multirow{3}{*}{252M} & Decoder & 58.8 & 43.9 & 69.0 & 37.7 & 39.9 & 63.3 & 53.1 \\
 & & CTC single-frame & 10.2 & 99.5 & 16.1 & 99.0 & 12.7 & 98.8 & 56.7 \\
 & & CTC segmental & 11.9 & 99.3 & 20.5 & 98.3 & 16.1 & 98.5 & 58.3 \\
\hline
\multirow{4}{*}{XLS-R + AFCM} & \multirow{4}{*}{318M} & CTC single-frame & 92.7 & 98.4 & 90.9 & 98.8 & 89.3 & 95.7 & 94.5 \\
 & & CTC segmental & 82.5 & 96.5 & \textbf{96.2} & 80.5 & 85.0 & 91.9 & 87.5 \\
 & & AF single-frame & \textbf{93.2} & 97.9 & 91.4 & \textbf{99.3} & \textbf{93.6} & 94.1 & \textbf{95.4} \\
 & & AF segmental & 2.8 & \textbf{100.0} & 1.9 & 99.6 & 1.7 & \textbf{100.0} & 50.8 \\
\hline
\end{tabular}

\vspace{0.5em}
\textit{(b) Japanese nasal: four-way classification}
\vspace{0.2em}

\begin{tabular}{llrccccc}
\hline
Model & \#Params & Method & {[m]} & {[n]} & {[\textipa{N}]} & {[\textipa{\textltailn}]} & Balanced Acc. \\
\midrule\midrule
\multirow{3}{*}{POWSM} & \multirow{3}{*}{252M} & Decoder & 53.7 & 68.1 & 63.9 & 1.0 & 46.7 \\
 & & CTC single-frame & 73.1 & 87.2 & 32.7 & 1.0 & 48.5 \\
 & & CTC segmental & 74.6 & 88.4 & 30.2 & 4.2 & 49.3 \\
\hline
\multirow{4}{*}{XLS-R + AFCM} & \multirow{4}{*}{318M} & CTC single-frame & 65.7 & 97.9 & 69.3 & 3.1 & 59.0 \\
 & & CTC segmental & 62.7 & \textbf{99.7} & 46.5 & 6.2 & 53.8 \\
 & & AF single-frame & 74.6 & 92.8 & 83.2 & 12.5 & 65.8 \\
 & & AF segmental & \textbf{86.6} & 71.6 & \textbf{93.6} & \textbf{38.5} & \textbf{72.6} \\
\hline
\end{tabular}
\end{table*}

\subsection{Experimental Results}

Table~\ref{tab:results} presents the per-IPA recall and balanced accuracy for each task.
POWSM yields poor accuracy on both tasks, indicating that training on G2P-labeled multilingual data is not sufficient for zero-shot phonetic discrimination.
The proposed method of using XLS-R + AFCM with CTC achieves 94.5\% on aspiration and 59.0\% on nasal classification, demonstrating the benefit of finetuning with PanPhon-tokenized labels.

\begin{figure}[t]
\centering
\includegraphics[width=0.89\columnwidth]{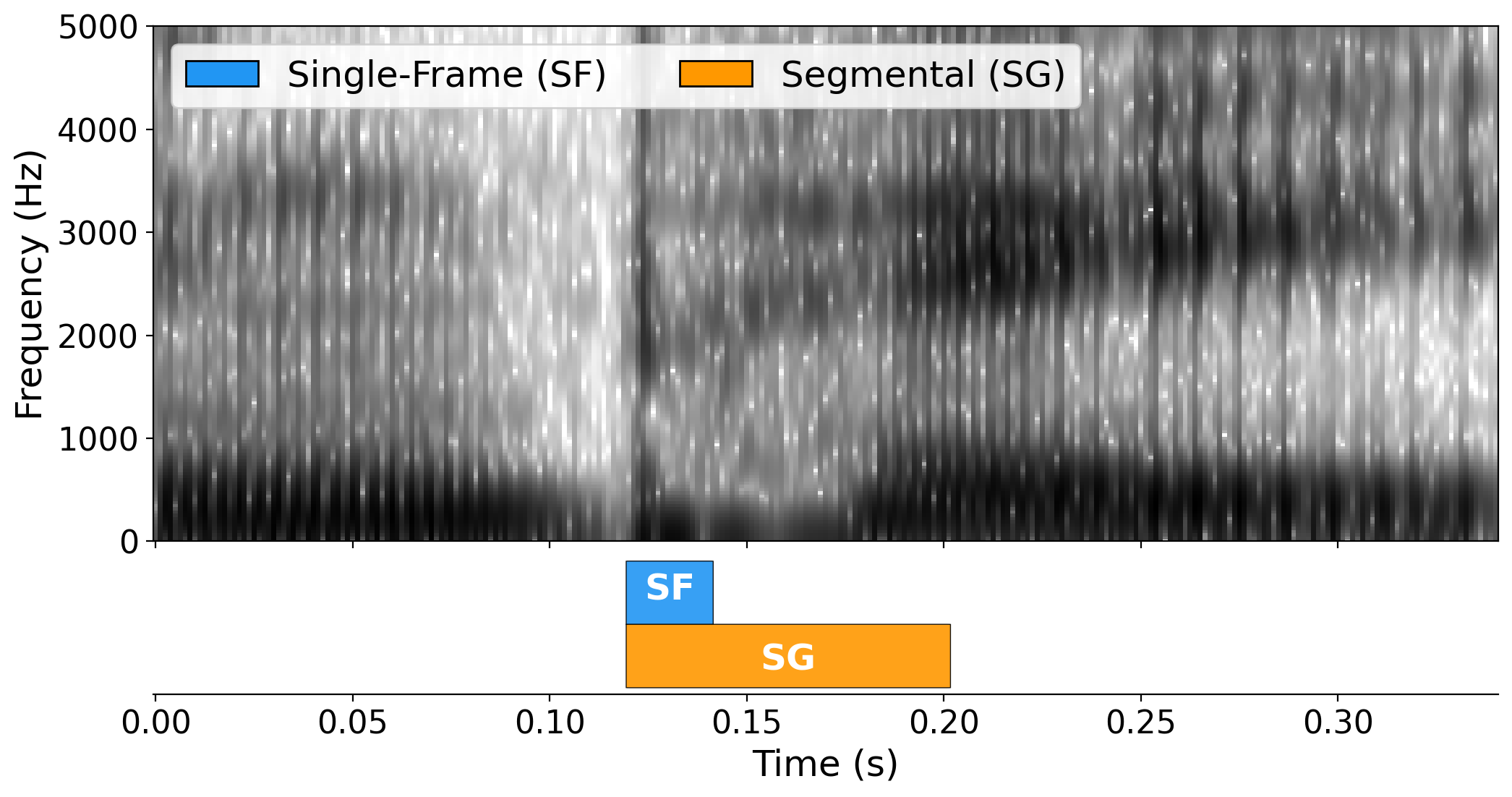}\\
\textit{(a) [p\textsuperscript{h}] (aspiration)}

\vspace{0.3em}
\includegraphics[width=0.89\columnwidth]{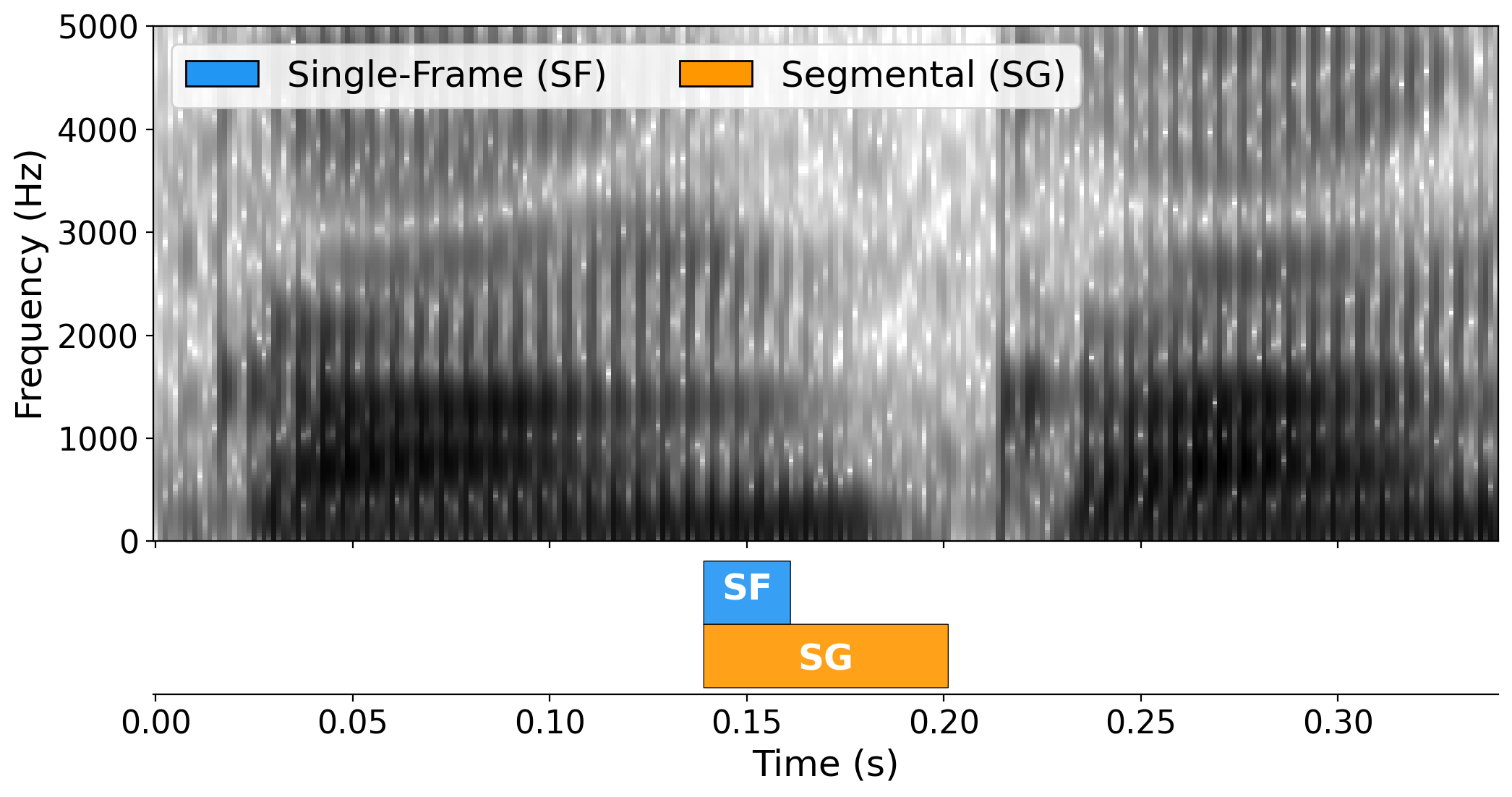}\\
\textit{(b) [\textipa{N}] (nasal)}
\caption{Spectrograms with single-frame (SF) and segmental (SG) windows. The aspiration burst in (a) is localized near S, while the nasal place cues in (b) persist across N.}
\label{fig:spectrogram}
\vspace{-1.5em}
\end{figure}

For aspiration (Table~\ref{tab:results}(a)), POWSM with CTC exhibits a strong bias toward unaspirated stops, recalling aspirated stops at only 10--20\% while unaspirated stops exceed 98\%.
XLS-R + AFCM with CTC single-frame resolves this bias, achieving over 89\% recall for all categories (balanced accuracy 94.5\%).
The single-frame AF method further improves to 95.4\%, with over 91\% recall across all categories.
However, segmental AF collapses to near-zero recall on aspirated stops ([p\textsuperscript{h}] 2.8\%, [t\textsuperscript{h}] 1.9\%, [k\textsuperscript{h}] 1.7\%), as the transient aspiration burst is diluted by averaging over the segmental window (Figure~\ref{fig:spectrogram}(a)).

For nasal classification (Table~\ref{tab:results}(b)), CTC-based methods achieve high recall on [m] and [n] but systematically misclassify [\textipa{\textltailn}].
The advantage of AF is pronounced on the rare categories: for [\textipa{\textltailn}], segmental AF achieves 38.5\% recall, whereas all non-AF methods remain below 7\%.
Since AF labels provide supervision for phonetic features shared across many phonemes, the model can learn acoustic-articulatory mappings even for phones that are rare in the training data.
For [\textipa{N}], segmental AF reaches 93.6\% compared to 69.3\% with CTC single-frame.
As shown in Figure~\ref{fig:spectrogram}(b), the place-of-articulation cues of [\textipa{N}] persist throughout the segment, enabling segmental aggregation to improve AF-based balanced accuracy from 65.8\% to 72.6\%.

Overall, these results highlight two findings. 
First, temporal aggregation using XLS-R + AFCM outperforms existing discrete CTC posteriors, suggesting that existing methods can further be improved.
Second, the effectiveness of the aggregation varies with the acoustic nature of the target distinction.
Single-frame classification is preferable for temporally localized cues such as aspiration, while segmental aggregation is beneficial for sustained cues such as nasal place of articulation. 
The choice can be determined by the characteristics of target phones.

\section{Conclusions}

This study focuses on zero-shot phonetic classification on Chinese aspiration and Japanese  moraic nasals.
Our results show that relying solely on discrete IPA tokens fails to discriminate unseen phones, while classification using continuous AF vectors substantially improves performance, particularly for underrepresented categories.
The optimal temporal aggregation varies with whether the target cue is transient or sustained.

These findings have implications beyond the two tasks studied here.
First, classifying conditioned allophones such as Japanese moraic nasals shows that AF-based methods can capture sub-phonemic variation.
Second, our framework can facilitate perceptual training of phones absent from the learner's L1 without native-speaker supervision, as it operates in a language-agnostic manner using articulatory representations.
Third, the proposed approach provides a step toward truly language-agnostic zero-shot ASR, where unseen phones can potentially be handled without language-specific adaptation.

\clearpage

\section{Acknowledgments}
This work was supported by JST NEXUS 251043539.

\section{Generative AI Use Disclosure}
The authors used generative AI tools for editing and polishing the manuscript. The authors take full responsibility for all content.

\bibliographystyle{IEEEtran}
\bibliography{mybib}

\end{document}